\documentclass[%
 pre,
 amsmath,amssymb,
 reprint,%
]{revtex4-1}

\usepackage{graphicx}
\usepackage{dcolumn}
\usepackage{bm}
\usepackage{xcolor}

\usepackage[utf8]{inputenc}
\usepackage[T1]{fontenc}
\usepackage{mathptmx}
\usepackage{etoolbox}
\usepackage{physics}
\usepackage{verbatim}

\makeatletter
\def\@email#1#2{%
 \endgroup
 \patchcmd{\titleblock@produce}
  {\frontmatter@RRAPformat}
  {\frontmatter@RRAPformat{\produce@RRAP{*#1\href{mailto:#2}{#2}}}\frontmatter@RRAPformat}
  {}{}
}%
\makeatother
\begin{document}


\title{Effect of environmental noise on charge diffusion in DNA: \\
Towards modeling its potential epigenetic impact in live processes}
\author{M. Rossini}
 \email{mirko.rossini@uni-ulm.de.}
\affiliation{Institute for Complex Quantum Systems, Ulm University, 89069 Ulm, Germany.
}%
\affiliation{Integrated Quantum Science and Technology (IQST), Germany.}

\author{O. Ammerpohl}
\affiliation{Institute of Human Genetics, Ulm University $\&$ Ulm University Medical Center, 89081 Ulm, Germany.}

\author{R. Siebert}
\affiliation{Institute of Human Genetics, Ulm University $\&$ Ulm University Medical Center}
\affiliation{Integrated Quantum Science and Technology (IQST), Germany.}

\author{J. Ankerhold}%
 \affiliation{Institute for Complex Quantum Systems, Ulm University, 89069 Ulm, Germany.
}%
\affiliation{Integrated Quantum Science and Technology (IQST), Germany.}

\date{\today}

\begin{abstract}
Charge diffusion through desoxyribonucleic acid (DNA) is a physico-chemical phenomenon that on the one hand is being explored for technological purposes, on the other hand is applied by nature for various informational processes in life. With regard to the latter, increasing experimental and theoretical evidence indicates that charge diffusion through DNA is involved in basic steps of DNA replication and repair, as well as regulation of gene expression via epigenetic mechanisms such as DNA methylation or DNA binding of proteins. From the physics point of view, DNA supports a metallic-like behavior with long-range charge mobility. Nevertheless, particularly considering a living environment, charge mobility in DNA needs to take into account omnipresent noise and disorder. Here, we analyze quantum diffusion of single charges along DNA-inspired two-dimensional tight-binding lattices in presence of different sources of intrinsic and environmental fluctuations. It is shown that double-strand lattices, parametrized according to atomistic calculations of DNA sequences,  offer a complex network of pathways between sites and may give rise to long-distance coherence phenomena. These effects strongly depend on carrier type (electrons, holes), the energetic profile of the lattice (composition) as well as the type of noise and disorder. Of particular interest are spatially correlated low-frequency fluctuations which may support coherent charge transfer over distances of a few sites.  Our results may trigger further experimental activities aiming at investigating charge mobility in DNA both in the native in-vivo context as well as on artificial platforms.
\end{abstract}


\maketitle

\section{Introduction}

The last decade has seen impressive activities to design structures on nanoscales based on atomic and solid-state platforms. Also living nature contains a series of substances and building blocks which have been subjected by molecular chemistry tools to assemble basic molecular units to larger structures and templates. One of the most prominent substances in this regard is desoxyribonucleic acid (DNA), the polymer composed of two complementary polynucleotide chains that represent the universal genetic code of living species. Its comparably simple building principle based on only four elementary bases (nucleotides) abbreviated with letters A, C, G, and T, gives rise to a plethora of two- and three-dimensional meta-structures. The Watson-Crick complementarity of DNA, where A pairs only with T and C only with G, leads to their assembly into a ladder structure which is then further folded into a double helix. The ladder consists of the two DNA strands each having a 5’ to 3’ orientation but placed in forward and reverse orientation through the complementary base pairing.  From the point of technological exploration of DNA, biochemistry is able to design sequences of variable lengths and tailored compositions out of these four bases. The binding-specific properties have made DNA also the ideal construction set to build tailored nano-structures, for example, in the field of DNA origami \cite{dna-origami2006,zhan2021}.  

Particular attention DNA has received in the context of molecular electronics where its conduction properties are explored e.g. when linked to metallic leads \cite{dekker2000,wang2018}. As typical for many organic molecular aggregates, long-range charge transfer has been observed, mediated through $\pi$-orbitals with a typical bandgap between the highest occupied molecular orbital (HOMO) and the lowest unoccupied level (LUMO) between 3-4 eV depending on the sequence of nucleobases. However, the DNA-core is coated by a negatively charged sugar-phosphate backbone and in biological contexts embedded in aqueous media\cite{Liepinsh1992,McDermott2017,DubouDijon2016} which may mask the nucleotide-based charge transfer. In the latter situation, in designed DNA oligomers in solution bound to optically excited donor molecules relatively long-lived charge mobility (up to ms) over relatively long distances (up to 2 nm) was observed in a series of experiments \cite{kohler1, kohler2, kohler3}. Other experiments seem to provide convincing evidence that charge mobility in DNA can be long-range (up to tens of nm) depending on perturbations in $\pi$-stacking of the bases. One suggested mechanism combines quantum mechanical delocalization over a few bases with hopping-like diffusion between those coherent islands.

Apparently, the challenge here lies in the intricate interplay of metallic-like behavior (coherent charge transfer) with disorder and noise induced by surrounding degrees of freedom. The goal of this work is to explore this interplay for single charge transfer along DNA double strands within a simplistic model in terms of tight-binding lattices. We consider shorter DNA oligomers where quantum coherence is expected to survive and, most importantly, parametrize the tight-binding structure with parameters originating from atomistic simulations \cite{hawke_tight-binding_2010}. 

The rationale behind this analysis is guided by the following question: To which extent are noise and disorder, particularly correlated disorder, supportive for long-range charge transfer through DNA also in an ambient living environment?  There is compelling evidence that nature explores charge fluctuations and transfer in processes such as DNA replication and repair, as well as in the epigenetic regulation of gene expression, where epigenetic regulation refers to stable alterations in gene expression potential that arise during development and cell proliferation \cite{Jaenisch2003}. For example, charge transfer within and across DNA strands is key for the function of some DNA binding cellular proteins including polymerases or sulfur-iron containing DNA repair enzymes \cite{OBrien2017}. Charge transfer can alter the higher-order DNA structure and consequently its accessibility to transcription factors. Similarly, charge localization/delocalization has the potential to alter binding efficiencies of DNA sequence- or structure recognizing proteins. Epigenetic modifying enzymes like DNA methyl transferases have been shown to interrupt charge transfer by flipping out bases from the nucleotide stack \cite{Isbel2022,Levo2014}. Thus, charge transfer through DNA constitutes an informational mechanism of living beings which does not alter the DNA sequence itself and, thus, can be regarded as epigenetic layer.

Double-stranded tight-binding models have been used previously \cite{lambropoulos_tight-binding_2019, Herb2024, albuquerque_dna-based_2014, Chargemigration,bittner_lattice_2006} to study the quantum dynamics of electronic excitations along nucleic acid sequences, however, not accounting for the impact of disorder, noise, and fluctuations induced by external (water, solvent, presence of counterions, electrodes, substrate, phonons), and internal (nuclei, base-pair sequence, geometry) factors omnipresent in an ambient living surrounding. In the following, we therefore analyze various sources, namely, thermal quantum fluctuations in the extreme limits of purely local and purely collective property as well as spatially uncorrelated and correlated disorder.  
To our knowledge, the current literature on charge dynamics on DNA-like models only sporadically considers the effects of any source of noise or disorder, whereas these phenomena are ubiquitous when realistically considering any biologically related model. We study simple sequences that are already relevant for epigenetic purposes: Hence, this work aims to align with a new line of research that combines the efforts of the physics, chemistry, and epigenetic communities, as proposed by Siebert et al\cite{Siebert2023}, with the potential to uncover the possible relevance of quantum phenomena for DNA dynamics during the life of cells.

The paper is organized as follows: We introduce the tight-binding model in Sec.\ref{sec:model}, and recall some relevant aspects of the Lindblad formalism in Sec.\ref{sec:lindblad_th}. while the following sections are devoted to showing the results of our simulations in the different scenarios we were interested in investigating, both analyzing unitary, dissipative (Lindblad-like) and disordered dynamics. Finally, in Sec.\ref{conc:sec} we provide an overview of the work described in this paper, together with an outlook of the refinements that can be made to further improve the analysis we have developed.

\begin{figure}
\includegraphics[width=8.8cm,keepaspectratio]{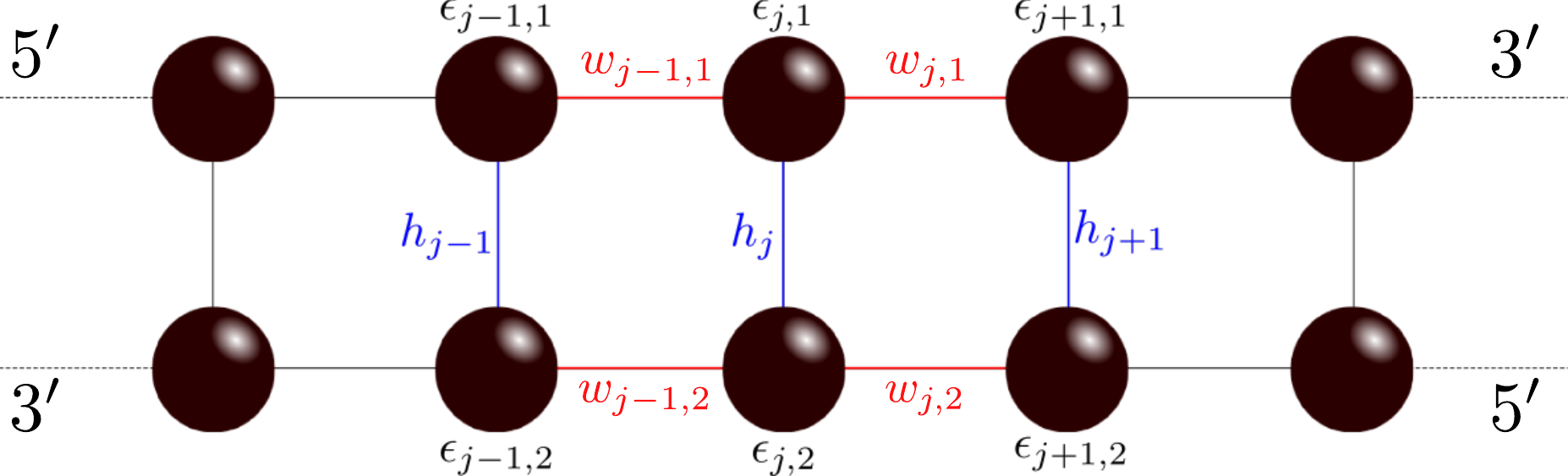}%
\caption{\label{dnamod}: Schematic representation of the TB Ladder Model (LM) used to mimic the structure of DNA.}%
\end{figure}


\section{The model \label{sec:model}}

We consider a model of the geometry and structure of DNA in form of a double-strand tight-binding ladder, called \textit{Ladder Model} (LM)  which has been introduced previously \cite{Chargemigration,bittner_lattice_2006} and is now well-established, see for example \cite{lambropoulos_tight-binding_2019}. The LM is indeed a minimal model that can address the influence of both base pairing and base stacking contributions in the energetics of the system. It consists of two tight-binding chains, denoted by the index $l\in L, \ \ L=\{1,2\}$, where
individual sites on each chain $l$ (nucleotides) are denoted by $1\leq j\leq n$ with  localized electronic (hole) states $|l, j\rangle$. The LM is parametrized by 
on-site energies of each base, $\epsilon_{j,l}$, intra-strand hopping integrals between successive bases, $w_{j,l}$, and inter-strand hopping integrals, $h_j$, see Fig. \ref{dnamod} for a schematic representation.
 Accordingly, the Hamiltonian reads
\begin{equation}
H_{LM}=H_{\rm loc}+H_{\rm tun}
\label{Hmodel}
\end{equation}
consisting of two parts: A local part describing on-site energies, $H_{\rm loc}$, and a non-local (tunneling) part describing interactions among different sites of the double chain, $H_{\rm tun}$, i.e.\ 
\begin{equation}
    H_{\rm loc} = \sum_{l\in L}\sum_{j=1}^N \epsilon_{j,l}\ket{j,l}\bra{j,l}
    \label{Hloc}
\end{equation}
and
\begin{eqnarray}
    H_{\rm tun}& = &\sum_{l\in L}\sum_{j=1}^{N-1}  w_{j,l}\ket{j,l}\bra{j+1,l} + h.c. \nonumber\\
    &&+ \sum_{j=1}^N h_{j}\ket{j,1}\bra{j,2} + h.c.\, .
    \label{Htun}
\end{eqnarray}

\begin{figure}
\hspace{-4.5mm}
\includegraphics[width=9cm]{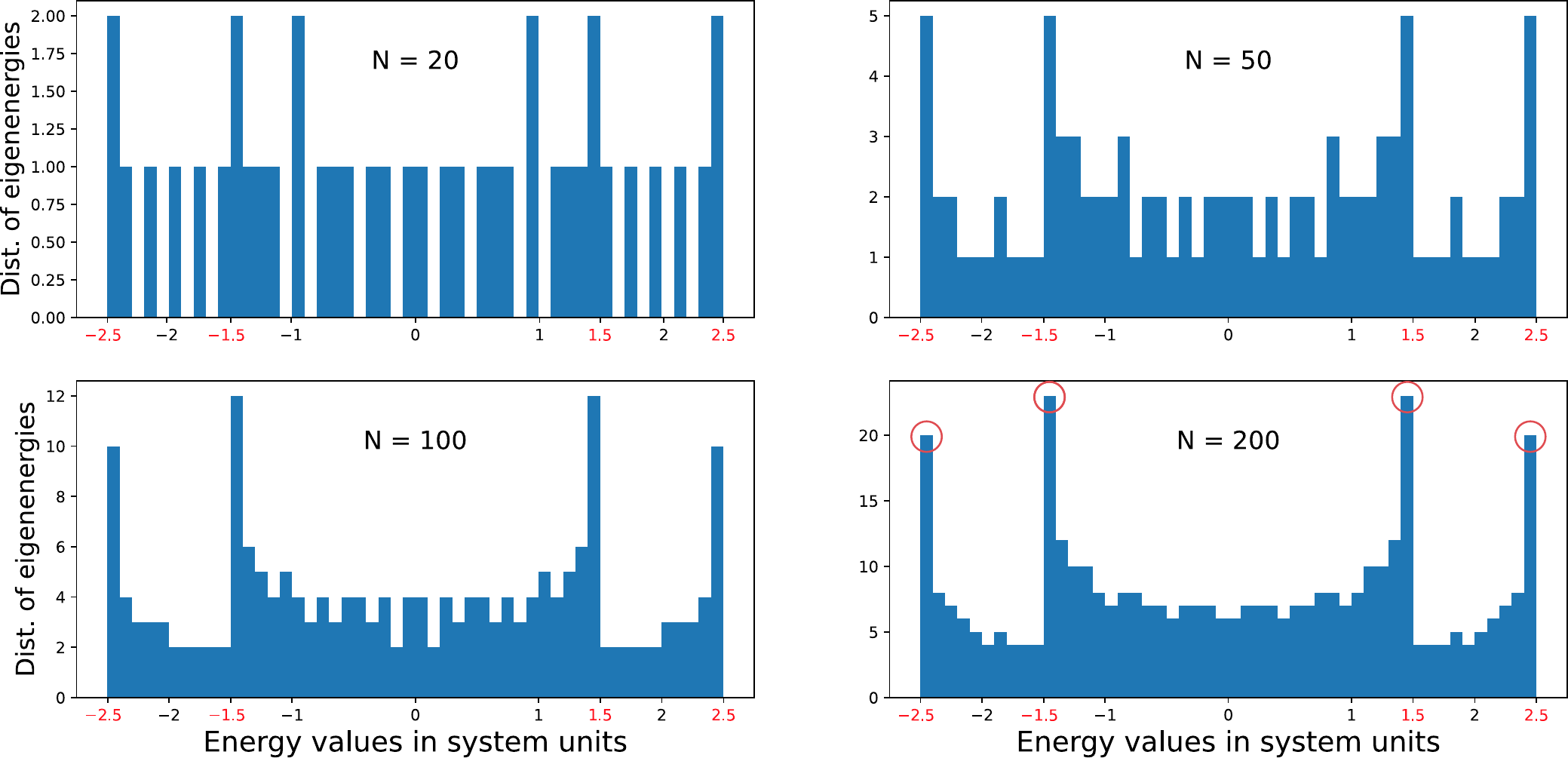}%
\vspace{-0.3cm}
\caption{\label{eigenen}: Distribution of eigenenergies (in units of $w$) of the system as a function of the total number of sites N. As can be shown analytically, all the values are ranging between -2.5 and 2.5.}%
\end{figure}

The LM allows for a straightforward solution in case of infinite chains and site-independent parameters $\epsilon$, $w$, $h$ throughout \cite{bittner_lattice_2006}. Even though this is a rather simplistic limit, it is nevertheless intriguing to briefly recall the main results. Assuming all parameters to be real-valued and positive and putting the lattice constant $a=1$, one finds energy bands
\begin{equation}
    E_{\pm}(k) = \epsilon + 2w \cos(k) \pm h\, , \ \ \ \ \ -\pi \leq k \leq\pi,
    \label{eigenenergies}
\end{equation}
with width $w$ and spacing $h$. They correspond to delocalized Bloch waves with quasi-momentum $-\pi\leq k\leq \pi$. 

For chains of finite length $N$ the quasi-momentum takes discrete values, i.e., 
\begin{equation*}
    k_n = \frac{2\pi}{N} \left(n-\frac{N}{2}\right)\, ,\ \ \ \ \ n = 0,...,N
\end{equation*}
and the expression in Eq.(\ref{eigenenergies}) remains valid upon substituting $k\to k_n$. Accordingly, molecular energy bands related to HOMO and LUMO cover a range from $E_{\rm min}=E_-(\pi)$ to $E_{\rm max}=E_+(0)$ and support coherent long-range charge transfer through the chain. For chains of growing length the expected density of states with pronounced maxima emerges. To illustrate the situation, we display in Fig.~\ref{eigenen}  an example with parameters $\epsilon = 0$, $w = 1$, and $h = 0.5$ and varying length $N$.

A realistic modelling has to include environmental degrees of freedom consisting e.g. local of vibrational modes as well as fluctuating modes of lower frequency stemming from the DNA backbone. Since a complete microscopic description is not feasible, one has to resort to effective description in terms of various kinds of noise models. In the sequel, we will start with broadband dynamical noise forces corresponding to local dissipation and then proceed with various kinds of nonlocal sources for decoherence.

\section{Dissipative quantum dynamics\label{sec:lindblad_th}} 

For broadband dynamical quantum noise, we follow the standard approach to describe the impact of thermal reservoirs on the quantum dynamics within the tight binding structure in case of weak interaction. The reasoning here is that charge transfer through DNA has been shown to relatively long-ranged and relatively fast compared to sequential processes, thus indicating relatively long-range coherence.
From now on we will consider the dynamics of a single charge/quasi-particle along the chain.

Due to weak coupling a perturbative treatment applies to derive an effective equation of motion for the reduced density operator $\rho(t)={\rm tr}_R\{W(t)\}$, where all environmental degrees of freedom are traced out from the total density operator $W$. More specifically, the total compound is described by a  Hamiltonian of the form $ H_{tot} = H_{S} + H_R + V$ with $H_S=H_{LM}$, $H_R$ modeling the reservoir modes, and $V$ capturing a bilinear coupling between LM and reservoir, i.e., $V= \sum_p A_p \, X_p$ with hermitian system operators $A_p$ and collective reservoir operators $X_p$. We note that the DNA-reservoir coupling can be very different in nature: For example, a uniform coupling, where all sites interact with the same reservoir, can be realized by setting  $ X\equiv X_i$ for all sites $i=1,\ldots, N$ with nonlocal system couplings $A_\nu$, while 
in a fully local coupling scheme site-dependent operators $X_i\neq X_j$ ($i\neq j$) appear with local operators $A_i$, while. All kinds of intermediate cases are possible as well, of course, but for the sake of clarity we will consider these limiting cases only. 

Under the assumption of factorizing initial states $W(t) = \rho(t) \otimes \rho_R(t)$ with reservoir equilibrium at inverse temperature $\beta=1/k_{\rm B} T$, i.e. $\rho_R=\exp(-\beta H_R)/Z_R$ [$Z_R={\rm tr}\exp(-\beta H_R)$],  and a time scale separation between the relaxation of the reduced system and the decay of equilibrium reservoir-reservoir correlations $\langle X_\nu (t)X_\mu(0)\rangle_\beta $, a Born-Markov treatment is valid. Together with a rotating wave approximation, this leads to   
\begin{equation}
	\dot{\rho}(t) = -i\left[H_S,\rho(t)\right] + \mathcal{L}[\rho(t)]
	\label{lindblad}
\end{equation}
with Lindblad superoperators of the form 
\begin{eqnarray}
	\mathcal{L}[\rho] &= &\sum_{\alpha, \beta}\sum_{\nu} \gamma_\nu(\omega_{\alpha \beta})\big[ A_\nu(\omega_{\alpha\beta})\rho(t)A_\nu^{\dagger}(\omega_{\alpha\beta}) - \nonumber\\  
	&&- \frac{1}{2} \{ A_\nu(\omega_{\alpha\beta})A^{\dagger}_\nu(\omega_{\alpha\beta}),\rho(t) \} \big]\, .
\end{eqnarray}
Here, the operators $A_\nu$ and $A^{\dagger}_\nu$, inducing transitions between eigenstates of the system Hamiltonian  $H_S\ket{\alpha} = E_\alpha\ket{\alpha}$, are represented in the form 
\begin{equation}
	A_\nu(\omega_{\alpha\beta}) = \ket{\alpha}\bra{\alpha}A_\nu\ket{\beta}\bra{\beta}
    \label{eigen-diss} 
\end{equation}
with transition frequencies $\omega_{\alpha\beta} = E_\alpha - E_\beta$. The interaction with the reservoirs is characterized by transition rates
\begin{equation}
	\gamma_\nu(\omega_{\alpha\beta}) = \int_{-\infty}^{\infty} dt\ {\rm e}^{i\omega_{\alpha\beta}t} \langle X_\nu(t) X_\nu(0) \rangle_\beta\, ,
	\label{diss-rates}
\end{equation}
with the assumption that cross-correlations between different reservoirs are absent.
Note that here any bath-induced renormalization of the transition frequencies is included in $H_S$. Given these rates, we can specify that the above time evolution equations are consistent if $\gamma(\omega_{\alpha\beta})\hbar/(k_{\rm B} T) \ll 1$ and $\gamma(\omega_{\alpha\beta})/\omega_{\alpha\beta}\ll 1$.

Even though Eq.(\ref{lindblad}) appears as an operator equation, implicit in its derivation is the representation of the density in the eigenstate basis of $H_S$. Making this explicit leads to two sets of equations, one for the diagonal elements (populations) $P_\alpha=\langle \alpha|\rho|\alpha\rangle$ which reads
\begin{eqnarray}
	\dot{P}_\alpha(t)& = &\sum_{\beta,\nu} \gamma_\nu(\omega_{\alpha\beta}) |\bra{\alpha}A_\nu(\omega_{\alpha\beta})\ket{\beta}|^2 P_\beta(t) -\nonumber \\ &&-\gamma_\nu(\omega_{\alpha\beta}) |\bra{\alpha}A_\nu(\omega_{\beta\alpha})\ket{\beta}|^2 P_\alpha(t)\, ,
	\label{diagonal}
\end{eqnarray}
and one for the off-diagonal elements (coherences)
\begin{equation}
    \dot{\rho}_{\alpha\beta}(t) = -\left( i\omega_{\alpha\beta} + \Gamma_{\alpha\beta} \right) \rho_{\alpha\beta}(t), \ \ \alpha \neq \beta\, .
    \label{offdiagonal}
\end{equation}
Here,  rates are given by
\begin{equation}
	\Gamma_{\alpha\beta} = \frac{\Gamma_{\alpha\alpha}+\Gamma_{\beta\beta}}{2} + \frac{\gamma(0)}{2}\, ,
	\label{eigen-rates}
\end{equation}
where $\Gamma_{\alpha\alpha} = \sum_{\beta \neq \alpha, \nu} \gamma_\nu(\omega_{\alpha\beta})$. 

Often it is very efficient and also very powerful to model reservoirs as quasi-continua of harmonic modes with spectral distribution $J(\omega)$. This is always possible if the fluctuation properties of a reservoir are Gaussian which is the natural situation for reservoirs with a macroscopic number of degrees of freedom. The impact of such reservoirs onto dedicated systems is then completely captured by second cumulants which in turn are determined by the spectral density $J(\omega)$ and temperature. Accordingly, the above transition rates can be expressed as 
\begin{equation}
	\gamma_\nu(\omega) = J_\nu(\omega)\left[ \coth(\tfrac{\omega\beta}{2}) - 1 \right]
	\label{rates-harmonic}
\end{equation}
with detailed balance property $\gamma_\nu(-\omega) =  J_\nu(\omega)\left[ \coth(\omega\beta/2) + 1 \right]$.

The second choice for the dissipators is to model the system-bath interaction as a coupling of each site with individual reservoirs (local dissipation). This can be done by making explicit the definition of the projection operators over each site as a linear combination of the system eigenstates, i.e.,
\begin{equation}
	A_{\nu,IJ} = \ket{I}\bra{I}A_\nu\ket{J}\bra{J} = \sum_{n,m}\alpha_{nm}\ket{\alpha_n}
   \bra{\alpha_n}A_\nu\ket{\alpha_m}\bra{\alpha_m}\, ,
	\label{local-diss}
\end{equation}
and then using these dissipators for the equations following  Eq. \ref{eigen-diss}. Note that this representation in terms of the eigenstate basis avoids a simpler and somewhat inconsistent representation in the local site basis. This latter representation is only valid if the tunneling terms in the total Hamiltonian (\ref{Hmodel}) are absent, i.e. in the trivial case. In the following, we will define this last choice of dissipators (see Eq.\ref{local-diss}) as \textit{Local Dissipators (LD)}, while we will refer to the first introduced set of dissipators (Eq.\ref{eigen-diss}) as \textit{Global Dissipators (GD)}.

Finally, we consider ohmic-type spectral bath distributions of the form
\begin{equation}
    J(\omega) = \eta\frac{\omega\Omega^2}{\Omega^2+\omega^2}
    \label{ohmic-spectral}
\end{equation}
with a coupling rate $\eta$ and a large cut-off frequency $\Omega$.

\section{DNA-parametrized lattices: Charge diffusion in presence of quantum noise}
\label{sec:dna-parameters}

To investigate charge dynamics on DNA strands, we make now use of specific parameter sets to mimic three different possible sequences of bases, namely, a degenerate sequence 5’-G-G-G-G-G-3’  (and complementary 5’-C-C-C-C-C-3’), a sequence 5’-G-C-G-C-G-3’ (complementary 5’-C-G-C-G-C-3’), and a sequence 5’-G-C-A-C-G-3’ (complementary 5’-C-G-T-G-C-3’). These short DNA-oligomers are chosen to cover both trivial sequences (dG - dC in the present work), which are already well studied in the literature, and more complex ones, whose resulting dynamics present interesting novel features and which are of potential relevance for epigenetic processes, particularly DNA methylation. DNA methylation usually occurs at the cytosine (i.e. C) in so-called CpG dinucleotides (where p denotes the binding phosphate). Thus, in addition to a poly-G stretch, a sequence with alternating CGs in both strands was chosen. This sequence constitutes a reduced representation of  a so-called “CpG Islands”. Such CpG islands are located in particular in promoter regions of many genes, particularly house-keeping genes. The cytosines in the CpG context can be methylated by sequence-specific DNA methyltransferases, a process during which charge transfer has been experimentally shown to be interrupted by flipping the cytosine to be modified out of the nucleotide stack. DNA methylation of CpG islands is frequently associated with the silencing of the expression of the linked gene \cite{Bird1986,Edwards2017,Ammerpohl2009}. For comparison, in the sequence 5’-CGTGC-3' after a first CpG, the sequence was interrupted by a T and inverted to a GpC, in which the G is not targeted by the DNA methyltransferases and, thus, not modified.

Corresponding tight-binding parameters are taken according to the work of Hawke \cite{hawke_tight-binding_2010} et al., which provides, to our current knowledge, the most comprehensive collection of data needed for our purposes based on Linear Combination of Atomic Orbitals (LCAO) calculations. 

Corresponding data are collected in Tab.\ref{DNAselfpar} and \ref{DNApar}.\footnote{We are aware that in the literature one can find different results for the parameters we are using in this model, which do not always agree with each other.  Moreover, we acknowledge the fact that in the epigenetic field, there are DNA sequences that are of the utmost importance for the purpose they serve, which are therefore good candidates for the analysis we want to perform.} We mention in passing that even with advanced DFT techniques only relatively short chains can be treated, typically no longer than 3-4 bases.

\begin{figure*}
\hspace*{-0.7cm}\includegraphics[width=18.5cm,keepaspectratio]{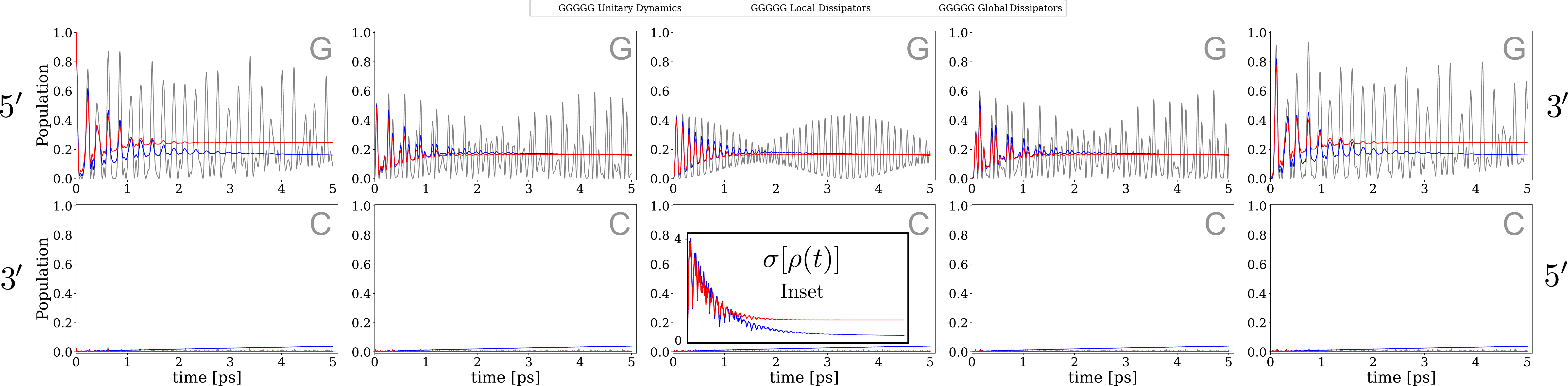}%
\caption{\label{GGGGGpop}: Population dynamics in the 5'-GGGGG-3' double-chain of an electron with unitary (grey line) and dissipative time evolution in the presence of a common bath (red line for global dissipation, blue line for local dissipation). In the bottom-central panel, an inset shows the time-evolution of the system coherence during the process.}%
\end{figure*}

\begin{figure*}
\hspace*{-0.7cm}\includegraphics[width=18.5cm,keepaspectratio]{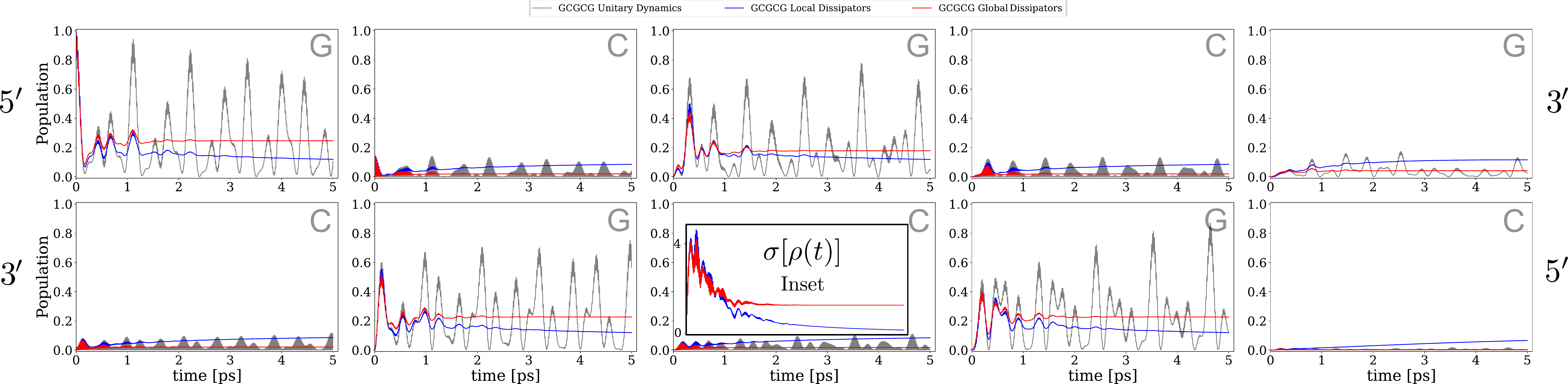}%
\caption{\label{GCGCGpop}: Population dynamics in the 5'-GCGCG-3' double-chain of an electron with unitary (grey line) and dissipative time evolution in the presence of a common bath (red line for global dissipation, blue line for local dissipation). In the bottom-central panel, an inset shows the time-evolution of the system coherence during the process.}%
\end{figure*}

\begin{figure*}
\hspace*{-0.7cm}\includegraphics[width=18.5cm,keepaspectratio]{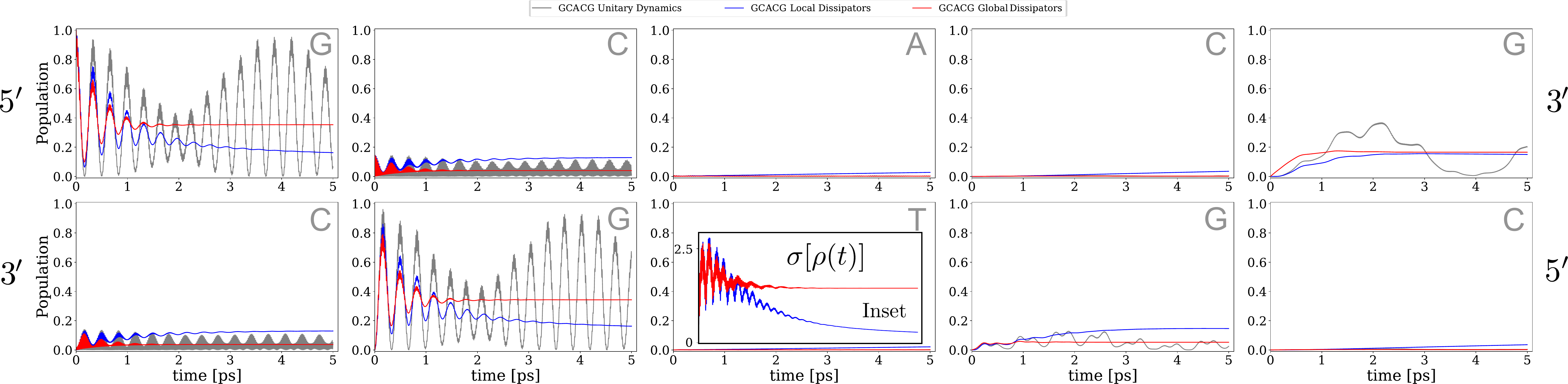}%
\caption{\label{GCACGpop}: Population dynamics in the 5'-GCACG-3' double-chain of an electron with unitary (grey line) and dissipative time evolution in the presence of a common bath (red line for global dissipation, blue line for local dissipation). In the bottom-central panel, an inset shows the time-evolution of the system coherence during the process.}%
\end{figure*}

After an evaluation of the dynamics resulting from such a setting in a closed environment, we introduce the two types of Lindblad operators (\textit{LD} and \textit{GD}) introduced in Sec.III to simulate decoherence for charge diffusion. All the resulting dynamics (unitary and open quantum) are plotted in Figs.\ref{GGGGGpop}, \ref{GCGCGpop} and \ref{GCACGpop}. The time-evolution of gross coherences for the two Lindblad-like evolutions are shown in the insets of the same figures mentioned above.

Figure~\ref{GGGGGpop} shows the quantum diffusion of a charge along a ladder consisting of five sites with G (top) - C (bottom) nucleotides. Initially, the charge is localized at the leftmost site on the upper strand. As expected, in the unitary case (grey line) the particle delocalizes relatively quickly along the upper G strand with almost no transfer to the complementary C strand though. This is in agreement with results found in the literature and is due to the large differences in on-site energies between the G and C sites, which are more than an order of magnitude larger than the energies associated with the overlap of the molecular orbitals of the two bases (see Tabs.\ref{DNAselfpar} and \ref{DNApar}). 

The dissipative dynamics is similar to the dissipative dynamics along a generic uniform chain, as expected. However, it is interesting to note the differences between different dissipators: The \textit{GD} partially preserves the coherence of the state even after for longer times, unlike the \textit{LD}. On the other hand, the charge dynamics induced by the \textit{LD} is able to populate sites that are not accessible in the unitary case and also in the \textit{GD} scenarios.

Figure~\ref{GCGCGpop} displays a more intriguing behavior of the quantum diffusion along the chain. Here, an alternating top chain  5’-G-C-G-C-G-3' with complementary 'letters' on the bottom side (5’-C-G-C-G-C-3') is studied. One clearly sees how, in the unitary dynamics, the topography of on-site and hopping energies force the charge to tunnel between guanine bases on one strand to the guanine bases on the opposite strand through cytosine bases. This process, known also as super-exchange, is a first example of non-trivial dynamics arising from tight-binding models parametrized according to the DNA energy topology. A similar phenomenon can be seen in Fig.\ref{GCACGpop} for the 5’-G-C-A-C-G-3' (5’-C-G-T-G-C-3') chain, where it is additionally favorable for the charge to tunnel even through the A-T base pair acting as a barrier in the middle of the chain. These data suggest that even in short and relatively simple DNA sequences, the quantum dynamics is prone to an interplay between quantum delocalization and back-scattering due to a non-trivial energy landscape along the chain.

A huge difference between the quantum dynamics in the \textit{LD} and \textit{GD} schemes appears along chains of the 5’-G-C-G-G-G-3' and 5’-G-C-A-C-G-3' types, as seen in Figs. \ref{GCGCGpop} and \ref{GCACGpop}. If reservoirs couple through local operators, we see complete destruction of coherence with a coherence measure $\sigma(t)$ quickly dropping below the value of $0.5$. In addition, we can observe population dynamics along the chain that differs substantially from the corresponding unitary evolution: While the unitary evolution is dominated by tunnelling through neighboring sites, thus populating almost only alternating G sites, \textit{LD} dynamics quickly starts to populate almost equally the entire chain, thus disrupting this structure of the charge distribution. In essence, thermal reservoirs kill coherent dynamics on a shorter time scale than the one required for the chain to show long-range quantum mobility.

The dynamics resulting from a \textit{GD} system-bath coupling scheme presents a very different behavior though. While we are still witnessing a gradual suppression of coherence, both achieve higher values of coherence in their dynamics, and their trends suggest that coherence will not collapse to zero for long  times. Alongside these results, we can also see major differences in the population dynamics. Indeed, while interferences and revivals along the chain are gradually washed out, only those specific sites that are also populated in the unitary dynamics discussed above get populated. Thus, the features arising from the topography of energy barriers (quantum tunneling) remain basically unchanged.

The main finding of this analysis can be summarized as follows: The way broadband reservoirs couple to DNA-type double-strand lattices with non-trivial energy landscape, namely, either nonlocal \textit{GD} or local \textit{LD}, has tremendous impact on the predicted charge diffusion process. It is not proven in the literature which coupling scheme is more suitable to mimic realistic DNA, in particular, there is no evidence to assume that DNA is predominantly locally coupled to  environmental degrees of freedom at the single base level. Now, our analysis suggests that a more subtle interplay between charge mobility along DNA lattices and quantum effects is required for a realistic description, in agreement with first experimental findings that report the existence of coherent 'islands' of up to 4-5 bases \cite{Genereux2009,Boon2002,Barton2019}.


\begin{table}
 \begin{tabular}{| l | l | l | l | p{3.5cm} |}
    \hline
    Adenine & Thymine & Guanine & Cytosine \\ \hline
     -4.4 eV &  -4.9 eV &  -4.5 eV &   -4.3 eV \\ 
    \hline
    \end{tabular}
 \caption{\label{DNAselfpar} On site energies of the DNA bases for the charge transfer model \cite{hawke_tight-binding_2010}}
\end{table}

\begin{table}
 \begin{tabular}{| l | l | l | l | p{3.5cm} |}
    \hline
    DNA bases & Tr. integrals & DNA bases & Tr. integrals  \\ \hline
    AA  & 16 \ \ \  & AT & 7\ \ \   \\ \hline
    AC & -3 & AG & 1  \\ \hline
    TT & -30 & TA & -7 \\ \hline
    TC & 22 & TG & -17 \\ \hline
    GG & 20 & GA & 30 \\ \hline
    GT & -32 & GC & 43 \\ \hline
    CC & -47 & CA & -12 \\ \hline
    CT & 63 & CG & 15 \\ \hline
    A-T & -9 & G-C & 16 \\
    \hline
    \end{tabular}
\caption{\label{DNApar} Intra-chain and inter-chain (last row) transfer integrals for the neighbours DNA bases and base pairs in the charge transfer model (in meV) \cite{hawke_tight-binding_2010}. Note that in general the values per base coupling do not commute, reflecting the nature of DNA of having one preferred direction along the strand.}
\end{table}

\begin{table}
 \begin{tabular}{| l | l | l | l | p{3.5cm} |}
    \hline
    Adenine & Thymine & Guanine & Cytosine \\ \hline
     -8.3 eV &  -9.0 eV &  -8.0 eV &   -8.8 eV \\ 
    \hline
    \end{tabular}
 \caption{\label{DNAselfparhole} On site energies of the DNA bases for the hole transfer model \cite{hawke_tight-binding_2010}}
\end{table}

\begin{table}
 \begin{tabular}{| l | l | l | l | p{3.5cm} |}
    \hline
    DNA bases & Tr. integrals & DNA bases & Tr. integrals  \\ \hline
    AA  & -8 \ \ \  & AT & 68\ \ \   \\ \hline
    AC & 68 & AG & -5  \\ \hline
    TT & -117 & TA & 26 \\ \hline
    TC & -86 & TG & 28 \\ \hline
    GG & -62 & GA & -79 \\ \hline
    GT & 73 & GC & 80 \\ \hline
    CC & -66 & CA & 5 \\ \hline
    CT & -107 & CG & -1 \\ \hline
    A-T & -12 & G-C & -12 \\
    \hline
    \end{tabular}
 \caption{\label{DNAparhole} Intra-chain and inter-chain (last row) transfer integrals for the neighbours DNA bases and base pairs in the hole transfer model (in meV) \cite{hawke_tight-binding_2010}. Note that in general the values per base coupling do not commute, reflecting the nature of DNA of having one preferred direction along the strand.}
\end{table}

\section{Comparing electron and hole dynamics}\label{el-hole}
The methods we have applied so far can also be used to study the quantum diffusion of holes (lack of an electron in the HOMO band) along DNA-mimicking structures. To do so, we need to choose a different set of parameters to mimik the behavior of this type of quasi-particles on double-chain sites. This set is given in Table \ref{DNAselfparhole} and \ref{DNAparhole} and is also taken from Hawke et al \cite{hawke_tight-binding_2010}. In particular, we can compare these results with those simulating the dynamics of an electronic charge along the same chain. Interestingly, the combined electron-hole dynamics may give rise to the formation of excitons which is discussed by Herb et al in \cite{Herb2024}. 

\begin{figure*}
\hspace*{-0.7cm}\includegraphics[width=18.5cm,keepaspectratio]{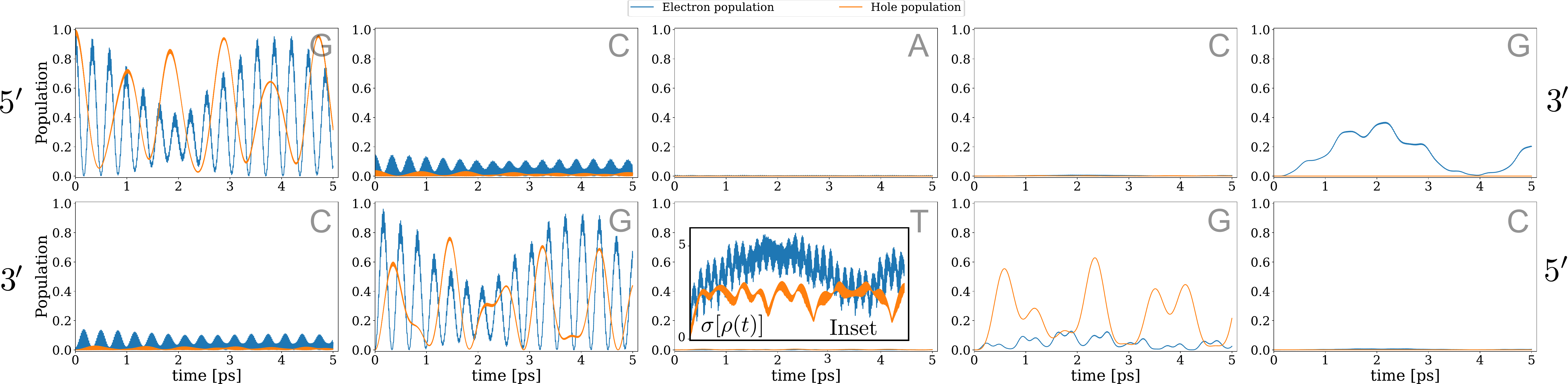}%
\caption{\label{GCACGUnitExcPop}: Population dynamics in the 5'-GCACG-3' double-chain for unitary time evolution and for both electron (blue) and hole (orange) transfer. In the bottom-central panel, an inset shows the time-evolution of the system coherence during the process for both electron and hole.}%
\end{figure*}

The results of the simulations are shown in Fig.\ref{GCACGUnitExcPop}, where for clarity only the results for the 5'-GCACG-3' sequence are shown, as these are the most relevant. We have already commented that in this scenario the electron is able to tunnel through the DNA sites to populate the guanine site on the right side of the chain. The hole dynamics, however, shows different tunneling characteristics. Namely, being able to populate the two guanine sites on the left side of the chain in a similar way to the electron dynamics, its behavior changes dramatically on the right side, where the hole can only occupy the guanine site to the right of the thymine site in the center of the chain. 

How electrons and holes populate the right-hand side of this chain can potentially be a strong source of resistance to exciton annihilation. In fact, the probability that an electron will relax back into the orbital from which it was excited depends strongly on the overlap of wavefunctions characterizing the electron and hole states at any point in time during the dynamics. Thus, if they are "trapped" far away from each other, the probability for excitonic annihilation is greatly reduced. As a possible consequence, long-lived excitation phenomena as observed in different experimental setups \cite{kohler1,kohler2,kohler3} could be explained in this way. These results suggest that investigating more thoroughly electron-hole relaxation could be critical to finally being able to link yet not completely understood experimental results with theoretical modelling. 

\begin{figure*}
\hspace*{-0.7cm}\includegraphics[width=18.5cm,keepaspectratio]{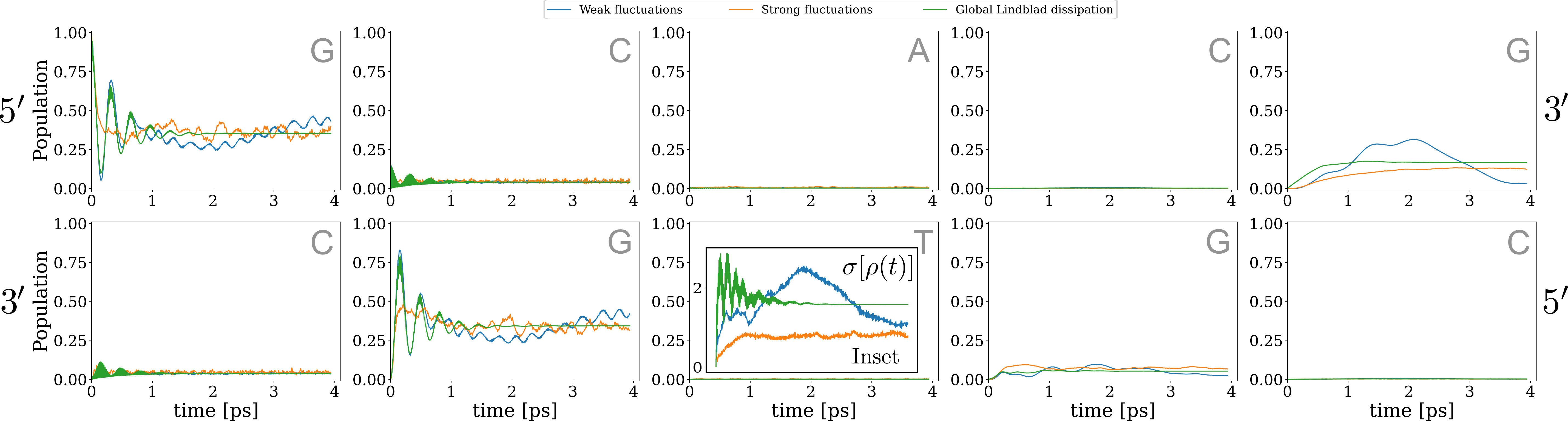}%
\caption{\label{GCACGedis}: Population dynamics in the 5'-GCACG-3' double-chain for static disorder and global Lindblad dissipation time evolution in electronic transfer. In the bottom-central panel, an inset shows the time-evolution of the system coherence during the process for all the dynamics plotted.}%
\end{figure*}

\begin{figure*}
\hspace*{-0.7cm}\includegraphics[width=18.5cm,keepaspectratio]{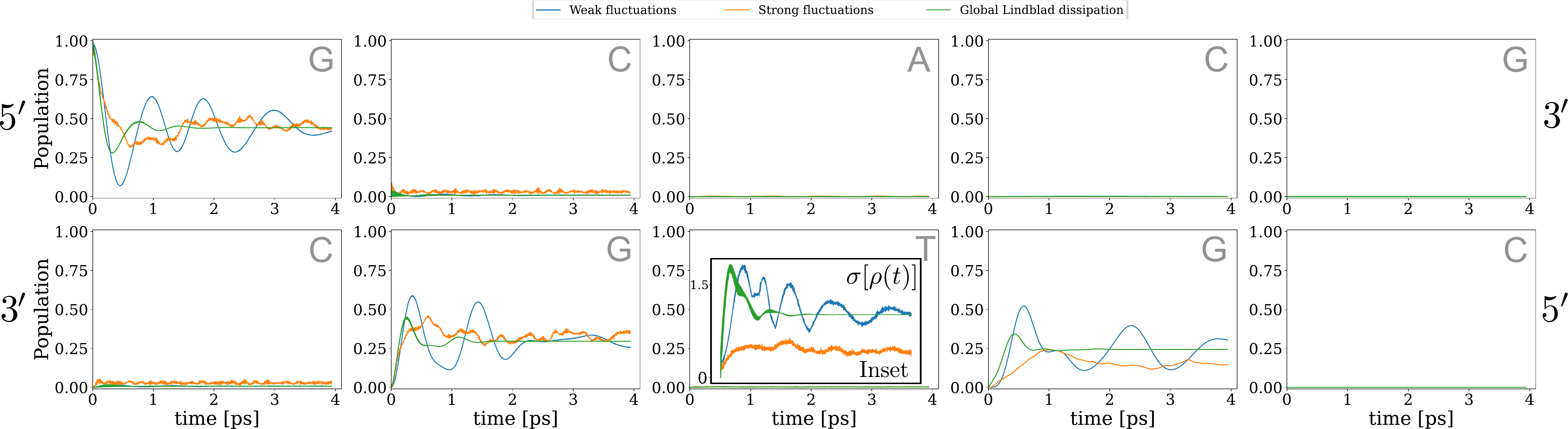}%
\caption{\label{GCACGhdis}: Population dynamics in the 5'-GCACG-3' double-chain for static disorder and global Lindblad dissipation time evolution in hole transfer. In the bottom-central panel, an inset shows the time-evolution of the system coherence during the process for all the dynamics plotted.}%
\end{figure*}

\section{Impact of static disorder}
Here we consider another source of decoherence with the tendency for localization of charges which can arise from static disorder in the local energy distribution of the DNA bases along the double strand. As suggested by Guti{\'{e}}rrez et al\cite{Gutirrez2009}, dynamic fluctuations in the energetics of the system can strongly affect how a particle propagates along the DNA chain, and this effect can also differ between electron and hole dynamics. We describe this by running several different dynamical simulations over the same DNA sequence, 5'-GCACG-3', where the site energies $\epsilon_{j, l}$ in (\ref{Hloc}) are replaced by $\epsilon_{i, l}+\xi_{i, l}$ with $\xi_{i, l}$ being a random variable sampled from a Gaussian distribution with a mean equal to $\epsilon_{i, l}$ and a variance to be set as an external parameter. We use two different values for the variance, the weaker one resembling energetic fluctuations typical of body temperature processes (about 27 meV), and the second, inspired by Guti{\'{e}}rrez et al\cite{Gutirrez2009} (360 meV), aiming to include even more sources of disorder, such as interactions with the backbone and solvent degrees of freedom. 

In Figs. \ref{GCACGedis} and \ref{GCACGhdis} we show the dynamics for an electron and a hole, respectively, both on the same DNA sequence, plotted together with their dynamics when in contact with a thermal bath modeled by global Lindblad operators. As for the plots above, the gross coherence dynamics for these simulations can be found in the bottom right panel's inset. It is interesting to note that the influence of static disorder on the dynamics of the particles is very similar to that of the global Lindblad operators: This is due to the fact that both sources of decoherence have the tendency to preserve the structure of eigenstates of the system, thus both mimicking low frequency fluctuations.

As a result, it seems evident that the main considerations we made about the relationship between electron and hole dynamics along the 5'-GCACG-3' strand hold true even in the presence of static disorder, both for small and large variations in the on-site energies, since this e-h separation effect appears stable under this noise source.

\section{Correlated noise sources}
Finally, we turn our attention to the effect of correlated fluctuations, as opposed to the uncorrelated ones presented in the previous section. Correlated fluctuations on local energies of the system are implemented inspired by an approach proposed by Liu et al\cite{Liu2016}. However, here we focus on the full quantum dynamics along the chain. First, we introduce a measure of the spatial correlation of the particle dynamics along the chain, defined by
\begin{equation}
    \chi(i)_\sigma^{d_0} = \frac{<P_0(t)P_i(t)>_t^{d_0, \sigma}}{<P_0(t)P_0(t)>_t^{d_0, \sigma}}\, .
    \label{eq:corr}
\end{equation}
Here, the symbol $<\cdot>_t^{d_0, \sigma}$ refers to the time average for a given choice of parameters $\sigma$ and $d_0$ (be described below), and $P_i(t)$ is the population over time at site $i$ of the system investigated, which in this case is the uniform 5'-GGGGG-3' double-strand. The reasoning to chose 5'-GGGGG-3' DNA double strands to demonstrate the impact of correlated low-frequency noise on the charge dynamics in a system with an already structurally high degree of charge mobility. The above measure allows us to track how correlated fluctuations of on-site energies affect the long-range charge propagation, and how such correlated noise may be beneficial/detrimental to support long-range coherence.

The noise profile for a given chain is assumed to be constant over the time span of a single run of the model's dynamics. For each run, a noise profile of degrading correlation strength with distance between sites is generated as follows: First, a value is extracted from a normal distribution with zero mean and variance $\sigma$ to be specified (it represents the strength of the energy fluctuations in the systems). This value is added to the local energy of the top left base of the system, representing its energetic disorder. The energy fluctuation of its neighboring base is then extracted from another Gaussian curve, now averaged around the value found for the first fluctuation and with a variance $\sigma^{*}$ derived as follows
\begin{equation}
    (\sigma^*)^2 = 2\sigma^2\left(1-C_{ij} \right) \, ,
    \label{eq:sigstar}
\end{equation}
where $C_{ij} = e^{-\frac{|i-j|}{d_0}}$ is the correlation strength between the two bases. Here $d_0$ represents the coherence length of the correlated noise (the longer $d_0$, the higher the correlation of the noise distribution in space, with the limit $\sigma^* \xrightarrow{} 0$ for $d_0 \xrightarrow{} \infty$). Thus, the further away two bases are, the larger the value of $\sigma^*$ will be, resulting in noise profiles that are more and more uncorrelated with distance. A more detailed derivation of such a result is given in the supplementary material of Liu et al\cite{Liu2016}, including the possibility of accounting for temporal correlations in the energy fluctuations, which is beyond the scope of this paper.

In a previous work, Kubař et al\cite{Kubar2008} showed how the average correlation strength between neighboring G sites on a DNA strand is $C^G_{i+1, i} \sim 0.7$, corresponding to a value of $d_0 \sim 2.8$. For this reason, in our investigation, we explore different choices of the parameter $d_0$ ranging from 1 to 5 and $d_0=10$ for neighboring bases. The intensity of the energy fluctuations, modeled by the parameter $\sigma$, is studied under different conditions, ranging from $\sigma = 0.1 \rm meV$ to $\sigma = 100 \rm meV$, in order to cover both weakly and strongly perturbed regimes with respect to the typical interaction energies of the system under study (see Tables \ref{DNApar} and \ref{DNAparhole}).

\begin{figure*}
\hspace*{-0.7cm}\includegraphics[width=18cm,keepaspectratio]{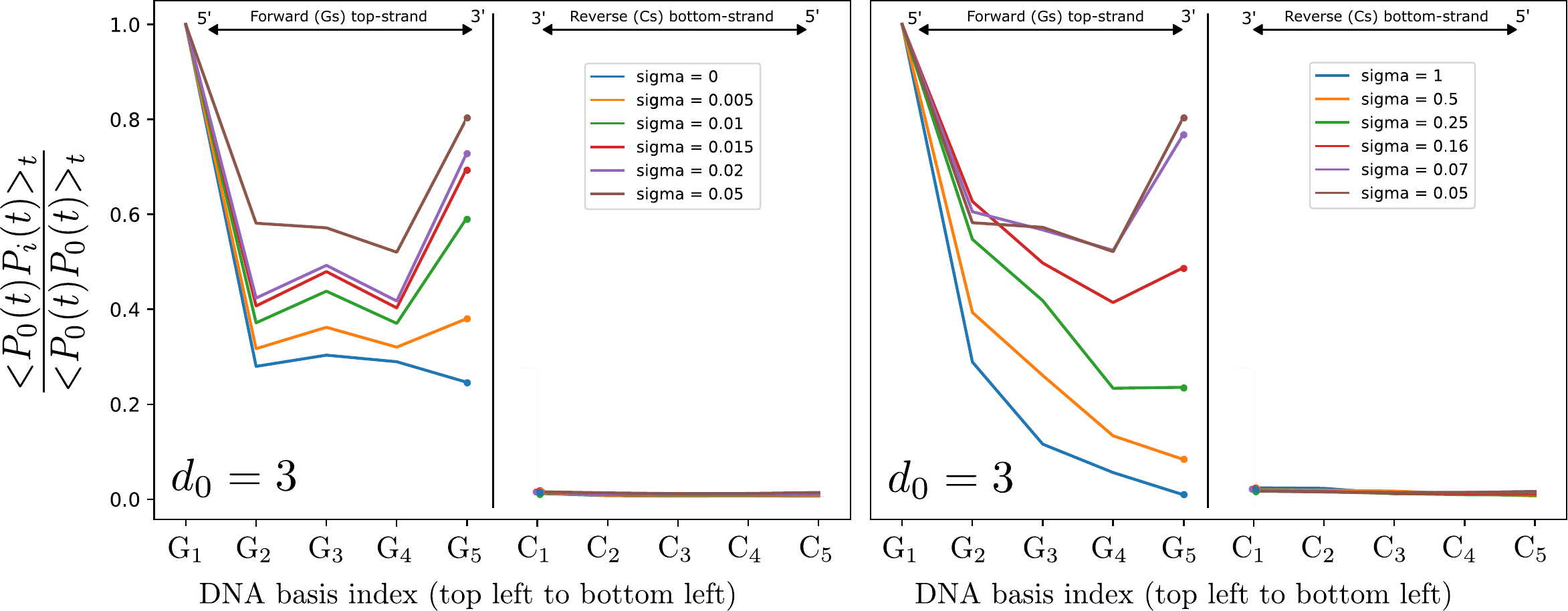}%
\caption{\label{GGGGGdyncorr}: Correlation measure for the time-averaged particle dynamics along each site (x-axis) of a double-stranded 5'-GGGGG-3' chain, for different values of the noise level ($\sigma$ parameter). The numerical labeling of the basis goes left to right for both strands (G$_1$ top left, C$_1$ bottom left)}%
\end{figure*}

\begin{figure*}
\hspace*{-0.7cm}\includegraphics[width=14cm,keepaspectratio]{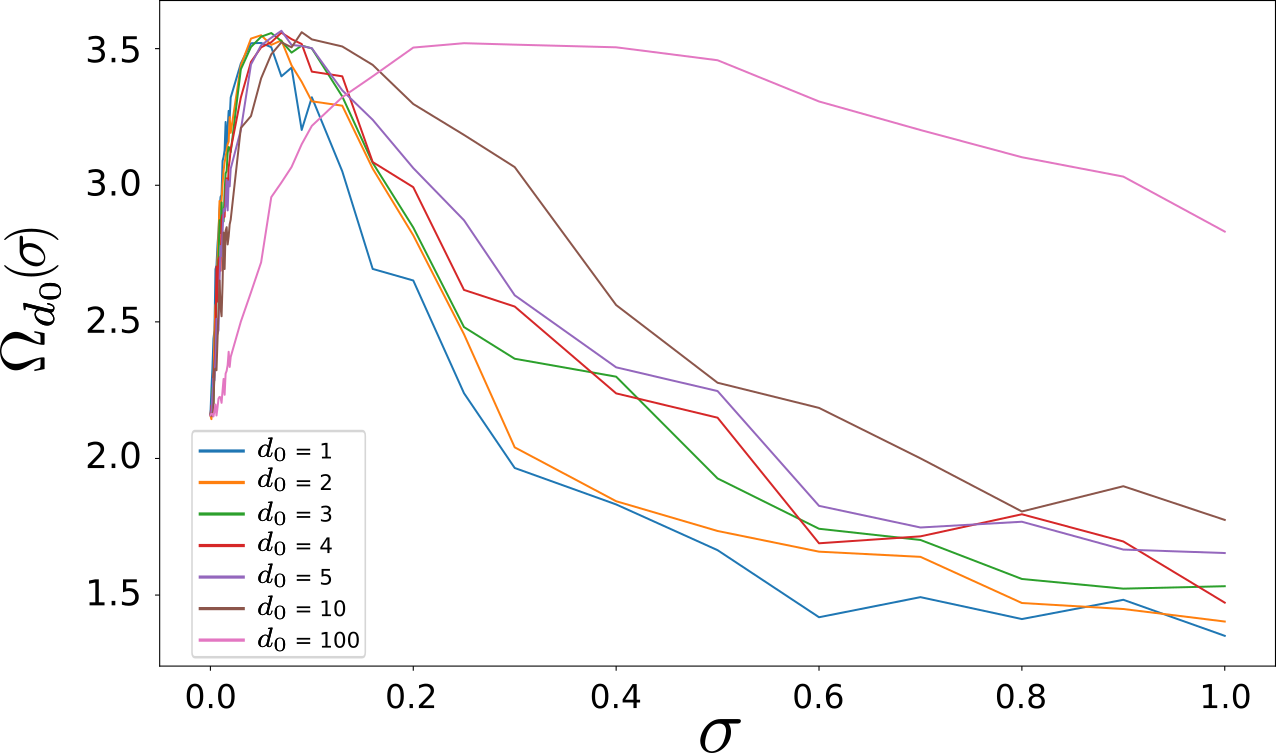}%
\caption{\label{sumGGGGGdyncorr}: Total correlation along a chain (sum over each site) for increasing $\sigma$ values (x-axis), for different correlation lengths. Smaller values of $\sigma$ are evaluated more finely (logarithmic selection of values).}%
\end{figure*}

Figure \ref{GGGGGdyncorr} shows, for a fixed parameter $d_0 = 3$ as a reference, the time-averaged correlation between the charge dynamics at each site of the DNA strand with respect to the dynamics at site $G_0$. The two plots show the same scenario for different choices of sigma. The left plot displays the degree of dynamical correlation of each site of the double chain for increasing values of the parameter $\sigma$, from $0$ to $0.05$. The right plot instead starts from the results for $\sigma = 0.05$ and increases this value up to the extreme choice of $\sigma = 1$. First, we can see that the correlation for the sites on the opposite strand of Cs is always zero. This reflects the fact that due to the energy topology the particle only delocalizes along the upper G-strand and never travels to the opposite chain. More interestingly, however, the data demonstrate that correlations along the G strand increase significantly with growing noise level, up to the value of $\sigma = 0.05$ (left plot in Fig. \ref{GGGGGdyncorr}), and only for even larger values begin to decrease, as would normally be expected from a measure of correlation versus noise level (right plot in Fig. \ref{GGGGGdyncorr}). The same can be observed for different values of $d_0$. 

To better investigate this aspect and to compare scenarios for different values of $d_0$, we considered the following quantity
\begin{equation}
    \Omega(\sigma)^{d_0} = \sum_{i}\chi(i)_\sigma^{d_0}
    \label{eq:sumcorr}
\end{equation}
which is plotted in Fig.\ref{sumGGGGGdyncorr}.
It is trivial to note that for $\sigma = 0$ all curves converge to a fixed value, since for zero fluctuations the contribution of the $d_0$ parameter is nullified. Moreover, along the rest of the curves, for larger values of $d_0$, the overall correlation $\Omega$ is generally higher for each value of $\sigma$ considered, reflecting the fact that for longer correlation lengths one expects a higher degree of correlation between the dynamics at different sites.

More importantly, however, these results illustrate that, for any value of $d_0$, the maximum degree of correlation {\em cannot} be found for $\sigma = 0$, as one would expect naively from the zero noise condition, but for a finite value of $\sigma$. This value depends on the correlation length of the systems under analysis. Hence, the analysis suggests that the presence of correlated static perturbations along the DNA does not necessarily disrupt coherent propagation but may even preserve it to some extent, in line with previous experimental findings \cite{Genereux2009,Boon2002,Barton2019}. However, this can only be the first step and further investigations toward understanding the true nature of the disorder effects on charge dynamics along DNA-like chains are required to fully understand their potential impact on DNA behavior.

\section{Conclusions}\label{conc:sec}
In this work, we investigated charge diffusion on two-dimensional DNA-inspired lattices in the presence of various noise mechanisms, from thermal reservoirs to spatially correlated disorder. The motivation for this study was triggered on the one hand by recent developments for long-range charge transfer through DNA and on the other hand by potential implications of charge delocalization/localization for DNA hemostasis and gene regulation, with a special focus on DNA modification by methylation.

The lattice model used here includes intra- and inter-strand site couplings and initial states prepared on one of the sites. For DNA lattices, parametrized according to atomistic simulations, in the absence of noise the energetic profile along complementary chains does lead to finite populations for spatially separated sites due to long-range tunneling related to long-range coherences due to proper $\pi$-stacking of bases. 

Weak coupling to quantum thermal reservoirs substantially modifies this picture, however, depending on the details of the coupling mechanism (local versus global). As a result, one observes that the assembling of specific sequences of nucleobases allows to control charge diffusion. While oscillatory patterns in the population dynamics die out rather quickly also for weak dissipation, interestingly, intra-strand coherences survive on relatively long times scales. For global quantum noise, coherences are present on much longer time scales which may be of use in set-ups, where molecular structures are placed in cavities. 

Electron and hole diffusion are quite different in DNA-like lattices, which can lead to complex exciton transfer in bare and weakly dissipative chains, as further investigated in \cite{Herb2024} for unitary dynamics. This phenomenon appears to survive in the presence of both Lindblad-like dissipation as well as for disorder noise, and we believe that further investigation may uncover interesting phenomena of potential biological relevance e.g. via epigenetic mechanisms.

The impact of disorder of  on-site energies of the system also revealed an interesting behaviour: Static disorder, both at low and high intensities, reveals dynamics similar to those induced by global dissipation which is somehow expected. In particular, when considering static disorder at room temperature energies, simulations suggest that coherences along the system are not suppressed on ultrafast time scales, but rather tend to have a relatively long-lived behaviour.

Even more interesting results have been found for  the effects of correlated disorder on the coherent propagation of an excess charge along the double strand. They suggest that a certain degree of disorder in the energy landscape of DNA could not only not hinder the coherent propagation of the particles, but even act as a support for it.

So far, we have investigated a rather simple model with the aim of providing general insight into the properties of quantum diffusion on DNA-like lattices when exposed to bath interaction effects. This may already trigger further studies on more specific situations where DNA is not in its native form as a carrier of genetic information, but rather as a tool to monitor quantum states in designed aggregates that are molecular or semiconductor-like in nature. Looking further into the future, however, one can imagine that these platforms could allow us to delve deeper into the charge transfer properties of specific DNA sequences including disease-associated mutations and the relationships between charge distributions and environmental degrees of freedom relevant to biological mechanisms affecting gene expression but not DNA sequence like epigenetic information.

\section{Acknowledgements}
We would like to thank Ciprian Paduariu, Björn Kubala, Jürgen Stockburger, Sima Karimi Farsijani and Emil Chteinberg for fruitful discussions. Moreover, we heartfully thank Dennis Herb for fruitful discussions and technical support during the project. Financial support of the Center for Integrated Quantum Science and Technology (IQST) and the BMBF through the project QCOMP in the Cluster4Future QSens is gratefully acknowledged.

\section{Data availability}
The data that supports the findings of this study are available within the article.

\bibliographystyle{ieeetr}
\bibliography{biblio.bib}

\end{document}